\documentclass[sigconf]{acmart}
\renewcommand\footnotetextcopyrightpermission[1]{} 
\usepackage{graphicx} 
\usepackage{tabularx}
\usepackage{changepage}
\usepackage{multicol}
\usepackage{caption}
\usepackage{subcaption}
\usepackage{tikz}
\usepackage{xcolor}
\usepackage{tcolorbox}
\usepackage{enumitem}
\usepackage{adjustbox}
\usepackage{float}
\usepackage{stfloats}

\begin{document}

\title{Driving with Guidance: Exploring the Trade-Off Between GPS Utility and Privacy Concerns Among Drivers}

\author{Yousef AlSaqabi}
\email{alsaqabi@usc.edu}
\affiliation{%
  \institution{%
    \begin{tabular}{@{}c@{}}
      The University of Southern California 
      \country{USA}
    \end{tabular}
    \\
    \begin{tabular}{@{}c@{}}
      Kuwait University
      \country{Kuwait}
    \end{tabular}
  }
}

\author{Souti Chattopadhyay}
\email{schattop@usc.edu}
\affiliation{%
  \institution{%
    \begin{tabular}{@{}c@{}}
      The University of Southern California 
      \country{USA}
    \end{tabular}
  }
}

\begin{abstract}

As the reliance on GPS technology for navigation grows, so does the ethical dilemma of balancing its indispensable utility with the escalating concerns over user privacy. This study investigates the trade-offs between GPS utility and privacy among drivers, using a mixed-method approach that includes a survey of 151 participants and 10 follow-up interviews. We examine usage patterns, feature preferences, and comfort levels with location tracking and destination prediction. Our findings demonstrate that users tend to overlook potential privacy risks in favor of the utility the technology provides. We also find that users do not mind sharing inaccurate or obfuscated location data as long as their frequently visited locations aren't identified, and their full driving routes can't be recreated. Based on our findings, we explore design opportunities for enhancing privacy and utility, including adaptive interfaces, personalized profiles, and technological innovations like blockchain.

\end{abstract}



\maketitle
\pagestyle{plain}

\section{Introduction}

Think back to the last time you used Global Positioning System (GPS) technology to navigate to an unfamiliar destination. Your experience was likely enhanced by a range of features designed to streamline your journey. With precise route guidance, you were confidently led along a path without having to second guess a turn. Real-time traffic information kept you informed of potential congestion and accidents on the road, helping you avoid delays and frustration. Alternate route suggestions ensured that you would not have to go through a specific street if you prefer not to. Additionally, trip time calculations allowed you to better plan your day by knowing an accurate estimate of when you would reach your destination. The seamless integration of GPS devices and smartphone applications into our daily lives has made navigation easier than ever before, transforming the way we travel~\cite{Skog}. 

However, as drivers become more reliant on GPS navigation systems, privacy concerns associated with using these systems have emerged as an important issue~\cite{Dey, Bi, Brush, Chitkara, Balebako, Cvrcek}. GPS devices and applications typically collect vast amounts of location data, which can be used to infer sensitive information about individuals, such as their habits, routines, and identity~\cite{Chorley, Gambs, Guha, Zhong}. As a result, there is growing unease among users regarding the potential misuse of their data by third parties, such as marketers, hackers, or even government agencies. Furthermore, the risk of data breaches and unauthorized access to personal information has intensified these concerns, raising questions about the ethical and legal aspects of GPS data collection and storage.

As GPS technology continues to evolve, understanding how drivers interact with these systems and addressing their feelings toward location data privacy is crucial for ensuring that the technology meets their needs, enhances their driving experience~\cite{Walker}, and fosters the development of technologies that strike a balance between offering valuable navigation services and protecting user privacy. While previous research has explored user preferences in older GPS technologies, these studies often do not account for the advancements and features present in modern systems. Moreover, although research exists on general attitudes toward location data privacy, these investigations have not specifically examined the unique context of GPS technology. Therefore, this paper is important because it bridges these gaps, providing a timely analysis of the trade-off between GPS utility and privacy concerns among drivers.

To address these concerns and to provide a comprehensive understanding of drivers' interactions with GPS systems, this study aims to answer the following research questions:

\begin{enumerate}
\item[\textbf{RQ1}:] How often do drivers use GPS?
\item[\textbf{RQ2}:] What preferences do drivers have when using GPS?
\item[\textbf{RQ3}:] How often does GPS suggest to drivers their preferred routes?
\item[\textbf{RQ4}:] How do drivers feel about their data privacy when using GPS?
\end{enumerate}

\begin{figure*}[ht]
    \centering
    \graphicspath{ {./Figures/} }
    \includegraphics[scale=0.5]{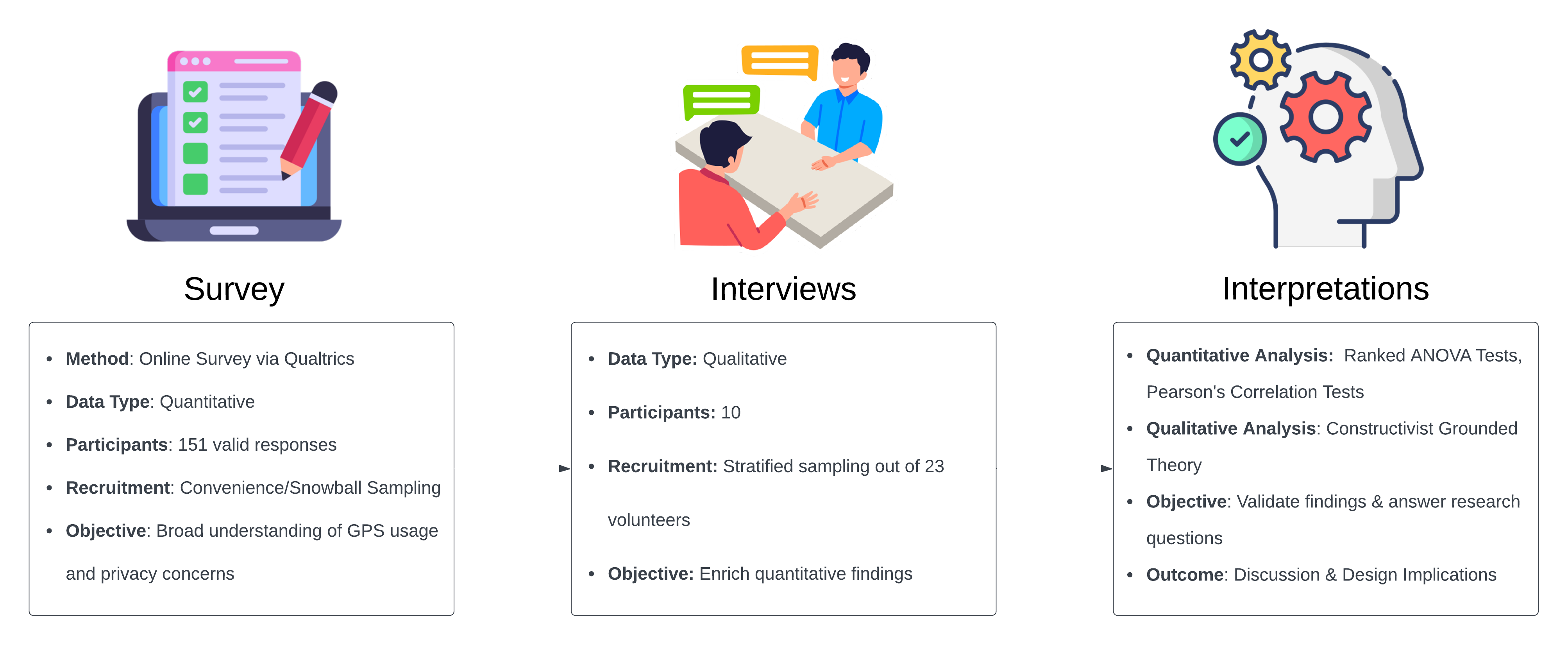}
    \captionsetup{font=small, labelfont=normalfont, textfont=normalfont}
    \caption{Flowchart illustrating the explanatory research design}
    \label{fig:method}
\end{figure*}

To explore these research questions, we conducted a study that adopts an explanatory mixed-methods approach as described in Fig.~\ref{fig:method}. The first phase involves a quantitative online survey to gather data on users' GPS usage, their preferences, and feelings towards their data privacy. The second phase involves qualitative interviews with a subset of survey respondents to delve deeper into the context of their choices and explore the reasoning behind their preferences and concerns.

This research paper is structured as follows: Section 2 provides a review of the literature on GPS navigation systems, divided into three parts: GPS usage and its impact on driving behavior and user preferences, the privacy concerns arising from sharing private data, and user studies focused on data privacy. Section 3 presents the methodology employed in the study, including the survey and interview design, sampling strategy, and data analysis techniques. Section 4 discusses the results of the survey and interviews, addressing the four research questions. Section 5 delves into the discussion, exploring design and technological solutions to enhance the balance between user privacy and utility. Section 6 discusses the study's threats to validity. Lastly, Section 7 concludes the paper, summarizing the main findings.

\section{Related Work}
\subsection{GPS Usage and Experience}

Many studies have explored the effects of GPS technology on driver behavior and preferences. Svahn~\cite{Svahn} found that drivers frequently use navigation systems when driving in a foreign traffic environment but only moderately use them in well-known environments. In line with this finding, Bonsall and Parry~\cite{Bonsall-P} found that most drivers travel along familiar roads to familiar destinations.

Drivers with high annual mileage regularly use navigation systems as decision support tools, supplementing their local knowledge~\cite{Svahn}. Familiar drivers demonstrate clear preferences among route alternatives and exhibit greater flexibility in their diversion behavior en-route~\cite{Lotan}. Bonsall~\cite{Bonsall} found a negative correlation between compliance and the level of familiarity, suggesting that personal preferences on routing become more important in well-known traffic environments. Familiar drivers prefer real-time information to make route choice decisions~\cite{May} and welcome information about scheduled disruptions and current network conditions~\cite{Bonsall-P}. 

Wallace and Streff~\cite{Wallace} also highlighted that drivers have different needs for route guidance information depending on the context. Krüger et al.~\cite{Kruger} discussed the importance of context by presenting a prototype navigation service spanning three different contexts: at home, in-car, and on foot.

Studies have also been conducted to compare the performances of different GPS navigation systems. Li et al.~\cite{Li} found that users perceived the routes generated by four different navigation apps as being of similar quality. However, Trapsilawati et al.~\cite{Trap} found that users had higher trust in Waze than Google Maps due to the higher degree of flexibility in information sharing between users, increasing system transparency. Users tend to use the navigation app they are familiar with, and prior experience strongly influences trust in the system~\cite{Trap}.

\subsection{Data Privacy Concerns}

Location privacy concerns have become increasingly prominent as an increasing number of location-based services continuously track users' locations, often without their knowledge~\cite{Dey, Bi, Brush, Chitkara, Balebako, Cvrcek}. This data is highly valuable, as it reveals personal information about users~\cite{Staiano}. Applications such as Google Assistant and Siri, as well as third parties like advertisers, can leverage this data to present marketing information or content tailored to users' interests and personal preferences~\cite{Guha}.

Numerous studies have demonstrated that developers and third parties can use machine learning techniques to infer users' personalities and demographics from location data shared on social networks~\cite{Chorley, Gambs, Guha}. Location information can reveal a wide range of personal and sensitive information about users~\cite{Chorley, Gambs, Zhong}. Users routinely share their location with various services and advertisers through mobile applications and libraries installed on their phones~\cite{Baron}.

As individuals, we produce a constant stream of information that can be used to observe and predict our spatial behavior, often without our consent or knowledge~\cite{goodchild}. Research by Abbas~\cite{Abbas} indicated that the amount of trust within a relationship determines people's willingness to share their location. However, those relationships are no longer easily identifiable, leading users to grant permissions they are asked for~\cite{Ricker}, consequently becoming willing cogs in the geosurveillance machine~\cite{swanlund}.

The incentives to opt-out or boycott are few, and frequent data leaks and the shadowy practices of companies like Uber and Facebook~\cite{degroot} strengthen the myth that everything is already in a database somewhere. As Thatcher~\cite{thatcher2017you} notes, "location" has become a product in itself, something that can be stored and exchanged~\cite{prudham2009commodification}. Location has transformed into a valuable commodity sought by large corporations as a powerful tool for shaping consumptive behavior~\cite{graham2022augmented}.

\subsection{User Studies on Data Privacy}

Several studies have investigated users' awareness of the privacy implications of permissions granted to applications and services, particularly access to location information. Golbeck et al.~\cite{golbeck2016user} and Chitkara et al.~\cite{Chitkara} both showed that users are largely unaware of these implications. Khalil and Connelly~\cite{khalil2006context}, Cvrcek et al.~\cite{Cvrcek}, Brush et al.~\cite{Brush}, and Toch et al.~\cite{toch2010empirical} have conducted various user studies to explore how individuals value their location data. Khalil and Connelly, for instance, studied the privacy preferences of their participants according to their willingness to disclose personal information, including location, to different entities such as family, friends, and work colleagues. They found that privacy is more desirable at home than at work, and participants are less likely to share their location information with colleagues and bosses than with significant others.

In a similar vein, Shih et al.~\cite{shih2015privacy} examined users' privacy preferences when sharing personal context (i.e., where they are, who they are with, and what they are doing) with third parties, specifically other mobile applications installed on their phones. The authors demonstrated that participants' decisions to share data are affected by the sensitivity of the data and the purpose for collecting it. They became more willing to disclose their context, even for private locations, when the usage of the data (purpose) is either clearly stated or missing. Shih et al. also showed that the purpose of the use of the location and its context is important to users and often not clearly indicated by the permission system of the phone; vague or non-explanatory purposes reminded users of the trade-off between privacy risk and the benefits of sharing data, making them unwilling to share it. Baron et al.~\cite{Baron} provided insights to set design guidelines for future privacy-preserving mobile systems and confirmed that people perceive different kinds of personal information with varying degrees of privacy preferences.



\section{Methodology}

\subsection{Survey}

The first phase of this study is a quantitative survey created through Qualtrics to gain a broad sense of drivers' GPS usage, preferences, and sentiments toward their location privacy. The target population for this survey is any person over the age of 18 who has a driver's license. Participants were recruited using a combination of convenience and snowball sampling.



\subsubsection{Study Design}

The online survey featured 20 questions, which were categorized as follows: 9 questions focused on participants' background and demographics, 4 questions addressed their GPS usage habits, 4 questions explored their preferences and experiences, 2 questions delved into their data privacy concerns, and a final optional question inquired about their interest in participating in a follow-up interview. Fig.~\ref{fig:survey} shows a subset of the questions asked in the survey.

\newcommand{\mybox}[1]{
    \begin{tcolorbox}[colback=gray!10!white,colframe=gray!50!black,arc=4mm,boxsep=0.3em,
    left=5pt,before={\vspace{-0.5em}\begin{minipage}[t]{\linewidth}},after=\end{minipage}\vspace{-1em}]
        \textbf{#1}
        \begin{itemize}
        \item[$\triangleright$] How often do you use GPS in known areas?
        \item[$\triangleright$] How often do you use GPS in unknown areas?
        \item[$\triangleright$] How often does GPS suggest your preferred route?
        \item[$\triangleright$] How comfortable are you with having your location tracked through GPS?
        \end{itemize}
    \end{tcolorbox}
}

\begin{figure}[!t]
  \centering
  \mybox{Survey Questions}
  \captionsetup{font=small, labelfont=normalfont, textfont=normalfont}
  \caption{A subset of the questions asked during the survey}
  \label{fig:survey}
\end{figure}

All of the questions were closed-ended, with three questions asking to rank a set of choices, while all the others were on a Likert scale. None of the questions were obligatory, and participants were allowed to drop out at any time. Depending on a participant’s answers, some questions were filtered and not presented to avoid asking unnecessary questions. While developing the survey, an external researcher reviewed the survey and provided feedback on the wording of the questions. The study was approved by our institution's IRB office. 

\subsubsection{Participants}

We collected a total of 151 valid responses, with participants taking an average time of 4.7 minutes to complete the survey. The mean age of the survey participants was 34.4 years old, with a standard deviation of 12.6, and 55.9\% of participants were male, with 44.1\% being female. Participants came from a variety of areas with different population densities: 37.3\% came from large cities, 20.4\% came from metropolitan areas, suburbs, and small cities each, and 1.4\% came from rural areas.

\subsection{Interview}

The second phase of the study involves qualitative interviews with a subset of survey respondents. From the previous survey, 23 participants volunteered to take part in a follow-up interview. Stratified sampling was used to select 10 interviewees from the pool of volunteers. The primary objective of these interviews is to enrich the quantitative findings by delving deeper into participants' experiences and the context behind their choices. By conducting these interviews, we aim to strengthen the validity of our findings and ultimately arrive at more insightful and well-informed conclusions.



\subsubsection{Study Design}

During the interviews, we initially inquired about participants' familiarity with GPS technology and their knowledge of data privacy, establishing a baseline understanding of their background on the topic. Subsequently, we asked participants about the frequency of their GPS usage while driving, the extent to which the GPS recommends their preferred routes, their willingness to change routes when prompted, and their general preferences in route selection. Finally, we delved into participants' perceptions of location privacy, their feelings towards their data being accessed and shared, and whether they believe avoiding such data exposure is possible. Although these questions are similar to the survey items, they provide crucial insights into the rationale and context behind participants' choices.

\subsubsection{Analysis}

In the data analysis of our study, the first author employed constructivist grounded theory~\cite{cgt} to qualitatively examine the interview responses. This approach allowed us to explore participants' experiences, meanings, and perspectives through an iterative and reflexive process. 


\section{Results}
\subsection{Survey}
\subsubsection{Quantitative Results}

\begin{figure}[!t]
    \centering
    \graphicspath{ {./Figures/} }
    \includegraphics[scale=0.19]{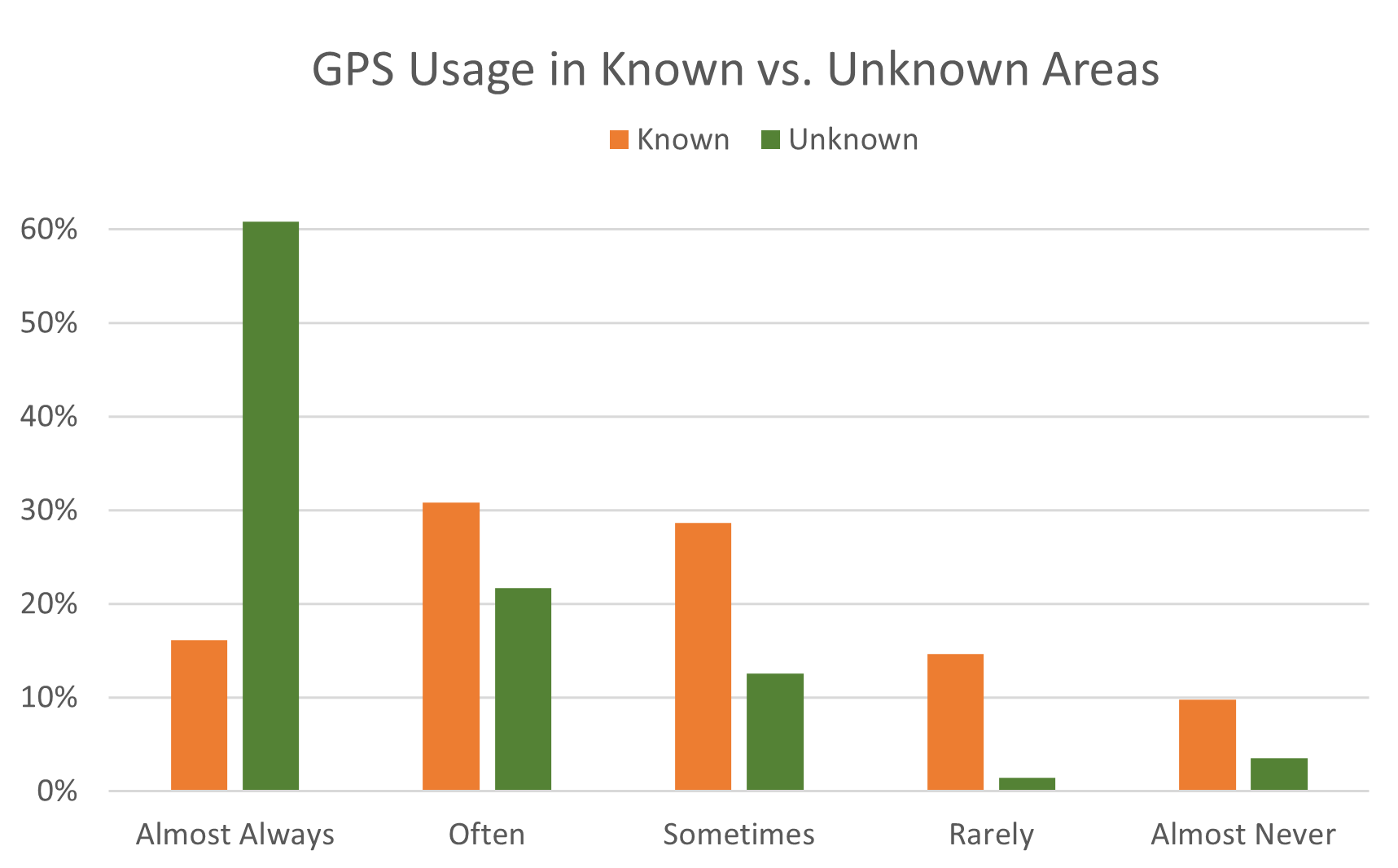}
    \captionsetup{font=small, labelfont=normalfont, textfont=normalfont}
    \caption{This chart shows how often GPS is used in known vs unknown Areas.}
    \label{fig:usage}
\end{figure}

\textit{RQ1: } Participants were asked how often they used GPS technology, specifically in known and unknown areas. Fig.~\ref{fig:usage} illustrated that while users do use GPS more often in unfamiliar areas, GPS is still frequently used even in familiar areas. 

\textit{RQ2: } Participants were then asked to rank the importance of five GPS functions on a scale of 1-5, with 1 indicating the most important: route guidance, real-time traffic information, address/places-of-interest search, trip time calculator, and geographical orientation/compass. The results revealed that route guidance was deemed the most important, with a mean score of 2.06, followed by trip time calculator (mean=2.65), real-time traffic information (mean=2.67), address/places-of-interest search (mean=2.75), and finally geographical orientation/compass (mean=4.87).

Following the ranking of GPS functions, participants were asked to rank three aspects of GPS path suggestions on a scale of 1-3, with 1 being the most important: fastest trip, least amount of traffic, and shortest distance. The results indicated that participants prioritized the fastest trip (mean=1.6), followed by the least amount of traffic (mean=2.08), and lastly, the shortest distance (mean=2.32).

\textit{RQ3: } When asked how often the GPS suggests to users their preferred path, most participants had positive path suggestions. 23.9\% responded almost always, 39.4\% often, 28.2\% sometimes, 5.6\% rarely, and 2.8\% almost never. When asked about the likelihood of switching to a new path suggested by the GPS while already driving, 20.3\% were extremely likely, 51.0\% somewhat likely, 15.4\% neutral, 12.6\% somewhat unlikely, and 0.7\% were extremely unlikely.  

\begin{figure*}[ht]
  \centering
  \begin{minipage}[c]{0.4\linewidth}
    \centering
    \includegraphics[width=\linewidth]{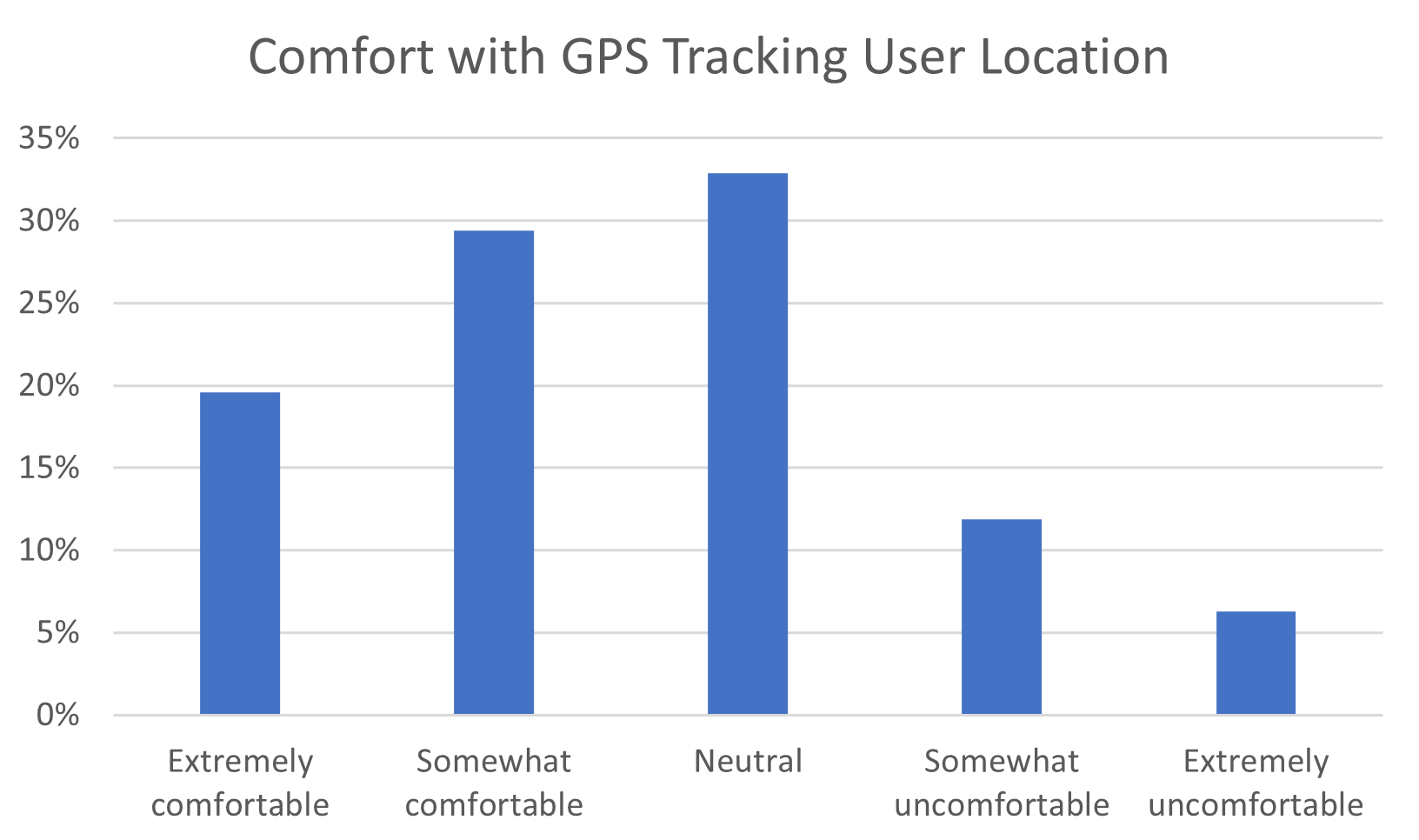}
  \end{minipage}
  \hspace{0.02\linewidth} 
  \begin{minipage}[c]{0.4\linewidth}
    \centering
    \includegraphics[width=\linewidth]{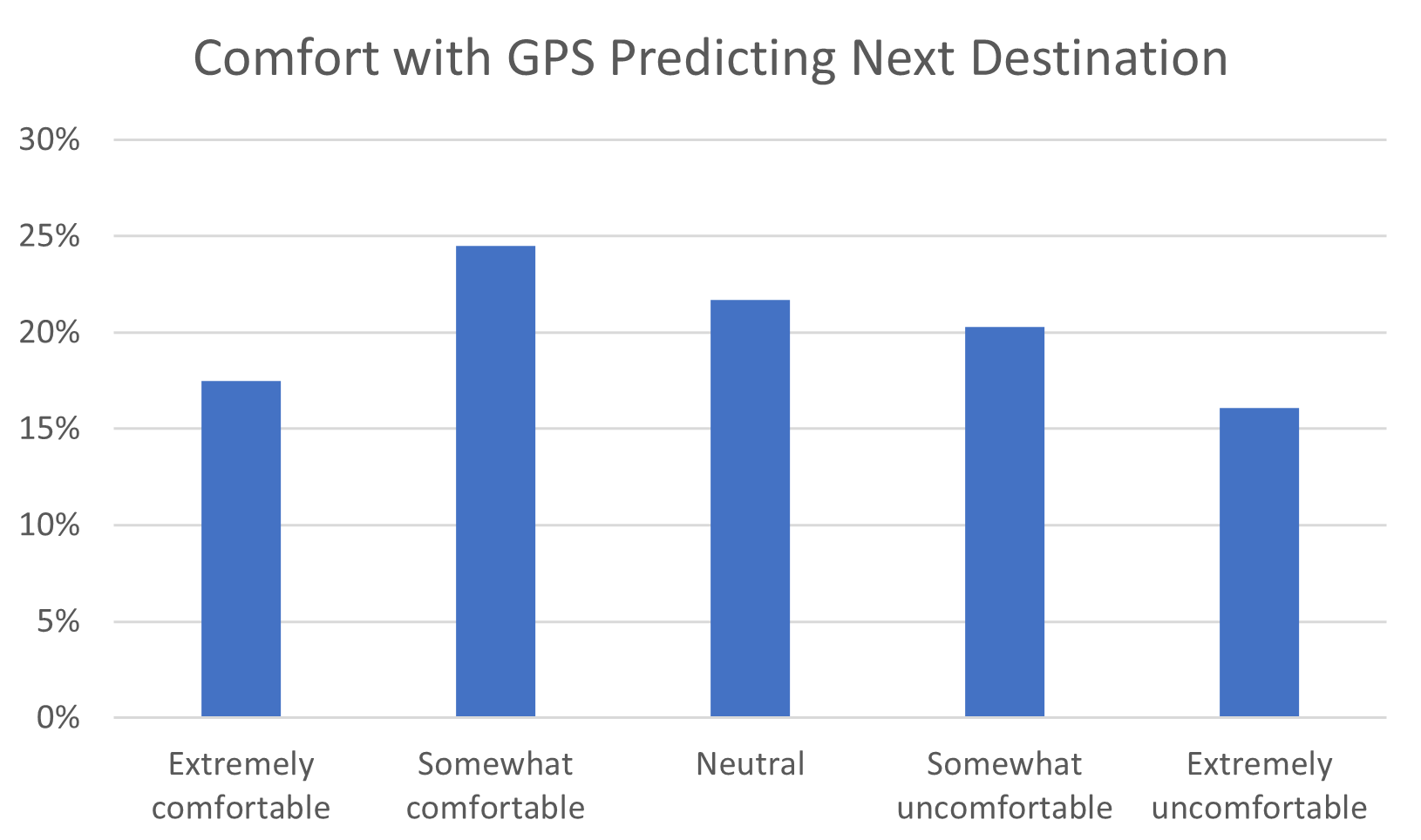}
  \end{minipage}
  \vspace{0.5\baselineskip} 
  \begin{minipage}[t]{0.4\linewidth}
    \captionsetup{font=small, labelfont=normalfont, textfont=normalfont}
    \caption{This bar chart shows how comfortable users are with having their location tracked through GPS.}
    \label{fig:privacy}
  \end{minipage}
  \hspace{0.02\linewidth} 
  \begin{minipage}[t]{0.4\linewidth}
    \captionsetup{font=small, labelfont=normalfont, textfont=normalfont}
    \caption{This bar chart shows how comfortable users would be with GPS predicting their next destination based on their previous driving habits.}
    \label{fig:prediction}
  \end{minipage}
\end{figure*}

\textit{RQ4: } Lastly, participants were asked how comfortable they are with having their location tracked through GPS, and how comfortable they would be if GPS could predict their next destination based on their previous driving habits. Fig.~\ref{fig:privacy} displayed that users were generally comfortable with having their location tracked, and Fig.~\ref{fig:prediction} revealed that users had mixed feelings towards the path prediction feature.

\subsubsection{Statistical Analysis}

In this subsection, we present the results of the statistical analysis performed on the quantitative survey data, using methods such as ranked ANOVA tests and Pearson's correlation tests.

Age was found to have a statistically significant impact on several aspects of GPS usage and preferences. A ranked ANOVA test revealed a significant relationship between age and GPS usage in unfamiliar environments (r(151) = .338, p = .011), with younger participants more likely to use GPS in unknown areas. Age was negatively correlated with the preference for the shortest distance as a factor in GPS path selection (r(146) = .25, p = .003), with younger participants more likely to choose the shortest distance as their least important preference. Conversely, age was positively correlated with the trip time calculator as an important GPS function (r(148) = .24, p = .005), with younger participants valuing this feature more. Lastly, age was positively correlated with the preference for the fastest trip as a factor in GPS path selection (r(146) = .21, p = .017), with younger participants more likely to prioritize this aspect. 

We also found a positive correlation between people's comfort with their location being tracked through GPS and their comfort with GPS predicting their next destination (p<0.00001). Moreover, a ranked ANOVA test found that there was a significant relationship between how often the GPS suggested participants' preferred routes and their comfort with having their location tracked (r(151) = .24, p = .003), suggesting that users may overlook potential privacy risks if they receive more utility from the technology. Another significant relationship was found between how often the GPS suggested users' preferred routes and their likelihood of switching to a new route while driving (r(151) = .32, p = .037). Users who experienced better initial path suggestions were more likely to trust and follow mid-route path suggestions, indicating that if the GPS can provide accurate initial path suggestions, users are more likely to trust the technology overall.

\subsection{Interviews}

The qualitative interviews aimed to provide a deeper understanding of participants' experiences and perceptions of GPS technology. Constructivist grounded theory was employed to analyze the interviews, yielding three main categories: GPS Usage and Experience, Trust and Reliance on GPS Technology, and Data Privacy Concerns. The first two categories add context to the initial three research questions, while the last category adds context to the fourth research question.


\subsubsection{GPS Usage and Experiences}

Participants demonstrated a lack of understanding of how GPS technology works, and many did not realize how their sensitive data could be maliciously leveraged without their consent. Participants reported frequent use of GPS technology while driving, regardless of their familiarity with the route: \textit{"I always use GPS, even if I am familiar with the area, to check traffic conditions."} (P2). 

Some participants emphasized using GPS technology for long-distance travel, not just for daily commuting. \textit{"For long drives, GPS is important not only for route guidance but for finding gas stations, rest stops, and food along the way."} (P6).

In some cases, participants experienced occasional signal loss or less accurate route suggestions, indicating potential areas for improvement in GPS performance. Many participants expressed frustration when encountering signal loss, especially in remote areas or when driving outside their country. This frustration was often due to signal loss happening when they didn't expect it, and the feeling of being lost associated with it. Anxiety was also commonly reported in these situations: \textit{"Not having signal would cause me emotional and mental distress, especially if I am not familiar with the place. It makes me anxious not knowing where to go."} (P5).

The usage of the share live location feature with friends and family was another common theme among participants. This feature allowed them to keep their loved ones informed about their whereabouts and estimated arrival times. Additionally, participants reported turning on location services for various phone applications, with Google Maps being the most popular navigation app. However, some participants also mentioned checking route suggestions from other apps like Waze before deciding on a path.

Preferred routes varied among participants, with some seeking the fastest route while others prioritized paths with fewer traffic lights or highways. The accuracy of GPS in suggesting preferred routes also differed, with some participants reporting high satisfaction with the suggested routes and others experiencing less accurate suggestions. Several participants mentioned double-checking alternate paths or using GPS primarily to estimate arrival times, even in familiar areas.

\subsubsection{Trust and Reliance on GPS Technology}

Generally, participants expressed trust in GPS technology and reliance on it for navigation. However, a balance between trusting the technology and using personal judgment or preferences emerged. Some participants mentioned double-checking alternate routes, comparing different GPS applications (e.g., Google Maps vs. Waze), and not always following suggested routes while driving, especially if they were already familiar with the area.

Participants shared their trust in the technology, as well as their concerns about becoming too dependent on it. \textit{"I appreciate GPS for easing my driving anxiety, but I worry about becoming overly reliant on it. It's important to keep some sense of direction without it."} (P9).

Another participant emphasized how GPS technology increased their confidence on the road: \textit{"I used to feel nervous driving in the city with all the intersections and busy roads. With GPS, I feel more secure knowing exactly where I need to go."} (P10).

Participants acknowledged the difficulties of avoiding GPS technology, given its utility in providing real-time traffic updates and efficient route planning: \textit{"it's very difficult to avoid, I could turn it off and not use it, but I would always be thinking If I could have gotten to my location sooner or avoided traffic"} (P2). This sentiment was echoed by other participants, who felt that the benefits of using GPS technology were too important to ignore.

The willingness to switch to a new suggested path while already driving varied among participants. Some were open to switching without much thought if it meant saving time, while others expressed concerns about the distraction it might cause or the confusion that may arise from changing routes mid-drive. Trust in personal driving instincts and sense of direction also played a role in this decision-making process: \textit{"I do not trust my driving instincts that well, so I follow the technology"} (P4).

\subsubsection{Data Privacy Concerns}

Participants exhibited varying levels of concern about sharing location data with GPS systems and third parties. While some felt uncomfortable sharing their data, others accepted it as a trade-off for the benefits provided by the technology, such as avoiding traffic jams or receiving real-time updates.

The level of comfort with data sharing varied among participants, with some expressing indifference or acceptance of the trade-offs: \textit{"Although it makes me feel uncomfortable, I feel like the benefits of using it outweigh my discomfort."} (P3). A distinction was made between sharing data with trusted entities, such as Google, and potentially malicious third parties, with the latter causing more concern: \textit{"I don't mind Google having access to my information since it probably has information on everyone, so I don't really feel targeted."} (P1).

Other participants shared their concerns around data privacy in the context of driving: \textit{"I'm not comfortable with my data being stored long-term. Sure, it can help predict traffic patterns, but the idea of someone knowing my daily commute is unsettling."} (P7).

In the case where third parties accessed participants' sensitive data, participants were asked about their thoughts on sharing inaccurate or incomplete location estimates to alleviate discomfort:\textit{"I would be comfortable with sharing just 25\% of my location data, as long as my complete driving routes couldn't be recreated"} (P8). Participants also expressed their discomfort with having their frequently visited locations identified.

Lastly, a participant suggested a novel idea for improving privacy while driving: \textit{"I wish there was a feature to use GPS anonymously, without storing my data, especially when I'm driving to personal places like home or family's house."} (P10). This reflects a desire for better control over personal data while using GPS services.

\section{Discussion}



One of the most pressing challenges in the realm of GPS navigation systems lies at the intersection of technology and human factors: finding the right balance between ensuring user privacy and providing a useful, personalized navigation experience. Our findings reveal that users may overlook potential privacy risks in favor of greater utility from the technology, while also highlighting the critical issues of signal loss and availability that directly impact the overall user experience. These multi-faceted challenges open up a plethora of design considerations for developers and Human-Computer Interaction (HCI) researchers, ranging from privacy risks to technological reliability.

\subsection{Designing for Privacy and Signal Reliability}

The varying comfort levels users have with location privacy offer a rich landscape for adaptive user interface design, emphasizing both ethical considerations and user-defined settings. Developers could create interfaces that are responsive to individual privacy preferences, such as implementing a "Privacy Mode" that obfuscates location details while maintaining essential navigation services. Visual cues, like a color-coded system, could further indicate the level of location data being shared or stored, enhancing user awareness and control. To empower users in making informed decisions about their data privacy, "Immediate Consent Notifications" could be introduced. These would appear when the app is about to perform a potentially privacy-invasive action, like sharing location data with third parties, allowing users the choice to opt-in or opt-out. This approach aligns with HCI principles of informed consent and user autonomy.

To address the concern of signal loss and interference, context-aware notifications could be integrated into the GPS interface. These notifications would alert users when they are entering areas with poor GPS signals and offer alternative routes or navigation methods. This feature could be particularly useful for users driving in unfamiliar areas, thereby improving the overall user experience and reducing anxiety related to signal loss.

\subsection{Personalized Navigation Experience}

Our study also revealed distinct user preferences for GPS features. By understanding users’ diverse needs and preferences, designers can create more effective and user-friendly GPS systems. This could involve offering customizable profiles or using machine learning algorithms to automatically adjust settings based on observed user behavior and feedback. 

For example, a "Time-Sensitive" profile could emphasize features that save time, while an "Eco-Conscious" profile might recommend fuel-efficient routes or those suitable for electric vehicles. A "Sightseeing" profile, on the other hand, would focus on scenic routes and points of interest for users more interested in the journey than the destination.


\subsection{Technological Solutions with HCI Implications}

Blockchain technology, as a decentralized solution, offers a promising avenue to address privacy concerns while still providing the benefits of GPS. By leveraging blockchain, location data could be stored and shared securely without relying on centralized data storage, thus enhancing privacy and control for users. From an HCI perspective, the challenge lies in designing interfaces that allow average users to interact with complex blockchain features without requiring a deep technical understanding. 

To further address signal reliance, machine learning algorithms could be employed to predict areas with poor GPS coverage. The HCI challenge here is to present these predictive analytics in a user-friendly manner, perhaps through intuitive visualizations or natural language explanations.

In terms of interaction methods, while multi-modal approaches like voice commands or haptic feedback may not directly solve the issue of signal loss, they can offer a more resilient and adaptable user experience under certain conditions. For instance, local caching of map data or predictive algorithms could allow these alternative interaction methods to serve as useful backups during periods of intermittent signal loss. Additionally, for planned trips, preloaded routes could enable basic navigation instructions to be provided through these alternative channels, even if the live map fails to update due to poor signal.

\section{Threats to Validity}

In this study, we have identified several potential threats to validity, which include internal, external, and construct validity concerns. We have taken measures to address these threats, but it is important to acknowledge them as limitations of our research.

\subsection{Internal Validity}

Survey participants may misinterpret the wording of questions or not respond as intended. To reduce this threat, we had an external researcher review the survey to clarify our wording. The use of convenience and snowball sampling in selecting survey participants may also pose a risk to internal validity, as this approach is neither randomized nor controlled. This sampling method might introduce biases in our sample, affecting the overall representativeness of our findings. Additionally, the recruitment process for follow-up interview participants, wherein individuals were asked if they would like to volunteer during the survey, may also threaten internal validity. The first author's acquaintances may be more inclined to volunteer, potentially leading to selection bias. 

\subsection{External Validity} 

In assessing the external validity of our study, it is essential to evaluate the extent to which our findings can be generalized beyond our specific participant sample to a broader population or different contexts. Although we collected responses from participants living in various areas, including metro areas, cities, suburbs, and rural locations, the generalizability of our findings is limited by the geographic distribution of our sample. With 60\% of the responses from Kuwait, 20\% from the USA, 14\% from Saudi Arabia, and the remaining from other countries, it is unclear how well our findings generalize to a global population. Differences in driving experiences, social norms, and country-specific laws may affect the applicability of our results.

\subsection{Construct Validity}

Our qualitative analysis relied on constructivist grounded theory, with the coding steps conducted by a single author. Although grounded theory is a suitable technique for single-author qualitative analysis, having multiple coders would help reduce potential bias.

\section{Conclusion}

In conclusion, this study explored the relationship between the utility and privacy concerns associated with using GPS technology. Our findings show that while GPS usage is higher in unfamiliar areas, GPS usage is still prevalent in familiar areas to assess traffic conditions and estimate arrival times. Users were found to value the route guidance feature the most out of all the GPS features but have mixed reviews on how well the GPS suggests their preferred routes. 

Regarding data privacy, our study showed that users had varying levels of comfort towards their location being shared and were more likely to be comfortable with having their location tracked if they already had positive experiences with path suggestions. This suggests that users may overlook potential privacy risks if they receive more utility from the technology. Participants have also expressed discomfort with having their full driving routes recreated and frequently visited locations identified. However, most participants did not mind sharing inaccurate or obfuscated data as long as the aforementioned conditions were met. 

In the discussion, we delve into the complex challenges of balancing user privacy and utility in GPS navigation systems, emphasizing the role of design and Human-Computer Interaction (HCI) principles. We suggest adaptive interfaces tailored to individual privacy preferences and introduce the concept of personalized navigation profiles to meet diverse user needs. We also discuss the importance of addressing signal loss through context-aware notifications and machine-learning algorithms. Lastly, we examine the potential of blockchain technology to bolster privacy and discuss the HCI challenges associated with making such advanced technologies user-friendly.

\newpage
\bibliographystyle{unsrt}
\bibliography{Bib.bib}

\end{document}